# Tc oscillations in multilayered cuprates superconductors

A. MESSAD

*Laboratoire de Physique, Lycée Olympe de Gouges, 205 rue de Brément, 93130 Noisy-Le-Sec, France.*

*When combining the BCS expression of Tc with a shear modulus constant model, we can demonstrate that Tc oscillates depending on the number of the cuprate planes and that it reaches its first maximum for three planes. This is in good agreement with experimental data and suggests that the electron- phonon coupling is essential to the superconductivity of the cuprates.*

**PACS numbers:** 74.62.-c; 74.72.-h

**Keywords:** cuprates; BCS; shear modulus; Tc oscillations.

## 1. INTRODUCTION

The critical temperature of the multilayered cuprates superconductors depends on the number of adjacent cuprate planes in the unit cell. If within a given cuprate family that is taken at the optimal doping, we vary the number n of the cuprate planes, we can observe that Tc follows a curve that looks like a dome whose maximum is at n = 3. This was experimentally well shown for the mercurocuprates family [1]. However, all experimentalists have pointed out the difficulty of elaborating and characterising compounds for which n is greater than 4 or 5 [2].
Recently, some experimentalists who used high quality materials have come to recognize that Tc is constant or hardly varies when n is superior to 5 or 6 [3]. Using the classic BCS expression of Tc together with a model of the shear modulus of a stacking of cuprate planes, we show that Tc oscillates depending on n and that the famous n = 3 arises naturally as being the localisation of the first maximum in the set of Tc's extremums.

## 2. Model and Application

According to BCS, Tc is proportional to the Debye temperature. The latter was given by Anderson [4]. So for Tc, we have

$$Tc \propto \left(\frac{h}{k_B}\right) \cdot \left(\frac{3qN_A}{4\pi\Omega_{mol}}\right)^{1/3} \cdot \left(\frac{G}{\rho}\right)^{1/2} \qquad (1)$$

and after some transformations

$$Tc \propto \frac{q^{1/3} \cdot G^{1/2}}{M^{1/2}} = f(n) \cdot G^{1/2} \qquad (2)$$

where $h$, $k_B$ and $N_A$ are respectively Planck, Boltzmann and Avogadro constants.
$\Omega_{mol}$ and $\rho$ are respectively the molecular volume and the density of the material.
G is the shear modulus of the crystal.
M is the molecular mass.
q is the number of atoms per molecule.

For a given cuprate family, the function f(n) is found using the chemical formula as shown by table 1.
Supposing that the cuprate planes behave like parallel springs, the shear modulus of the $p^{th}$ plane is given, as in metals [5], by

$$G_p \propto \frac{\sin(k_0 a_0 p)}{k_0 a_0 p} \qquad (3)$$

The summation over n planes gives the global shear modulus

$$G \propto \int_0^n \frac{\sin(k_0 a_0 p)}{p} dp \propto si(k_0 a_0 n) \qquad (4)$$

where $si$ is the sine integral function [5], $a_0$ is the interplane distance and $k_0$ is a vector of the reciprocal space. We suppose that $k_0 = 1/a_0$.
Thus the global shear modulus is simply given by

$$G \propto si(n) \qquad (5)$$

Finally if we suppose that the electron-phonon coupling parameter is a constant for an entire family [7], we obtain for Tc the following expression

$$Tc \propto f(n) \cdot [si(n)]^{1/2} \qquad (6)$$

In most cuprate families, f(n) hardly varies and thus Tc is reduced to

$$Tc \propto [si(n)]^{1/2} \qquad (7)$$

The plot of Tc versus n (fig.1) shows a set of extrema which are localised at π, 2π, 3π, …. The first one is a maximum and appears at exactly n = π (i.e. 3 with 4.7% error).
When applying equation (6) under the following form

$$Tc = A \cdot f(n) \cdot [si(K \cdot n)]^{1/2} \qquad (8)$$

to the Hg-cuprates and $Tl_2$-cuprates, we obtain the results of fig.2 and fig.3. A and K are the fitting parameters. The latter is found equal to unity as previously supposed.

## 3. Conclusion

According to this study, the critical temperature of the multilayered cuprates oscillates depending on the number of the cuprate planes. This fact is a consequence of a double coupling: first of an electron-phonon one in each cuprate plane and secondly of a mechanical one between the cuprate planes. An unexpected result which appears concerns f (n), this formula shows that using greater number of lighter atoms can achieve a higher Tc.
More precise experimental measurements will be necessary to confirm the existence of all the extrema.

## References

[1] K. H. Bennemann and J. B. Ketterson, *The Physics of Superconductors,* vol.1 (Springer 2003) p.423.
[2] P. J. Ford and G. A. Saunders, *The Rise of the Superconductors,* (CRC Press 2005) p. 60.
[3] A. Iyo, Y. Tanaka, Y. Kodama, H. Kito, K. Tokiwa, T. Watanabe, *Physica C* 445- 448 (2006) 17.
[4] C. P. Poole, *Handbook of Superconductivity,* (Academic Press 2000) p. 570 and 252.
[5] C. Kittel, *Introduction to Solid State Physics,* (8[th] edition, Wiley) p. 103
[6] N. N. Lebedev, *Special Functions & their Applications,* (Dover 1972) p. 33.
[7] to be published.

*Acknowledgements* **:** I wish to thank Mrs F. Pillier for providing me with the most part of the bibliography. I am grateful to Miss C. Martin for her help in writing the english version.

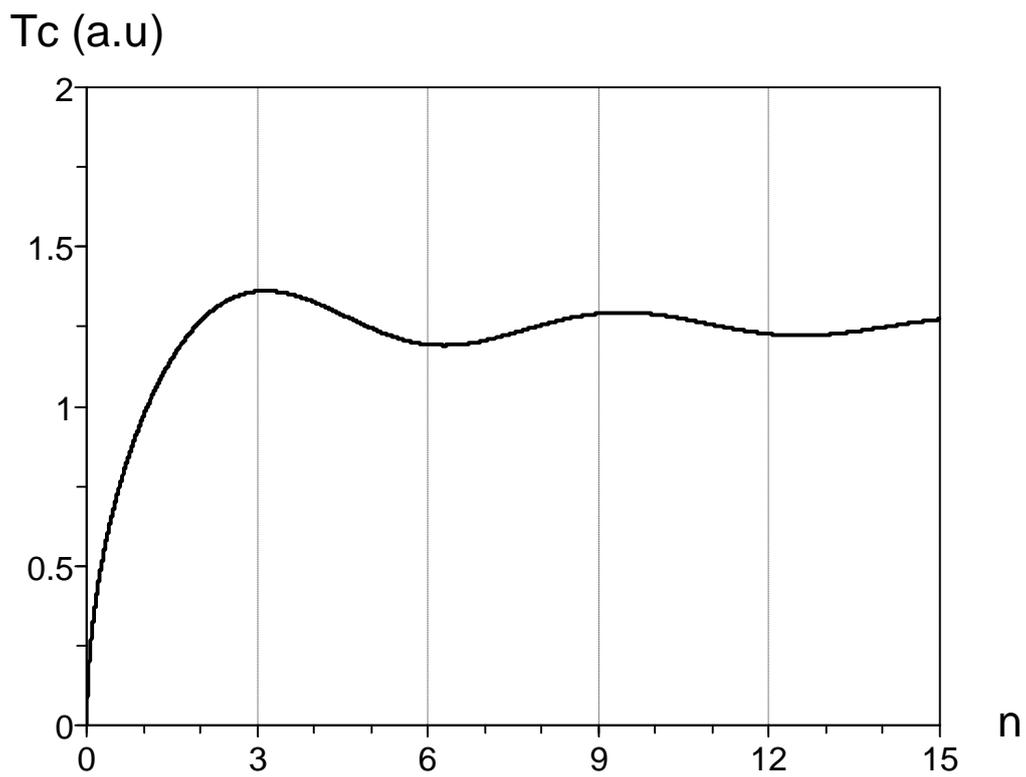

Figure 1: Plot of Tc *vs.* n from Eq.(7).

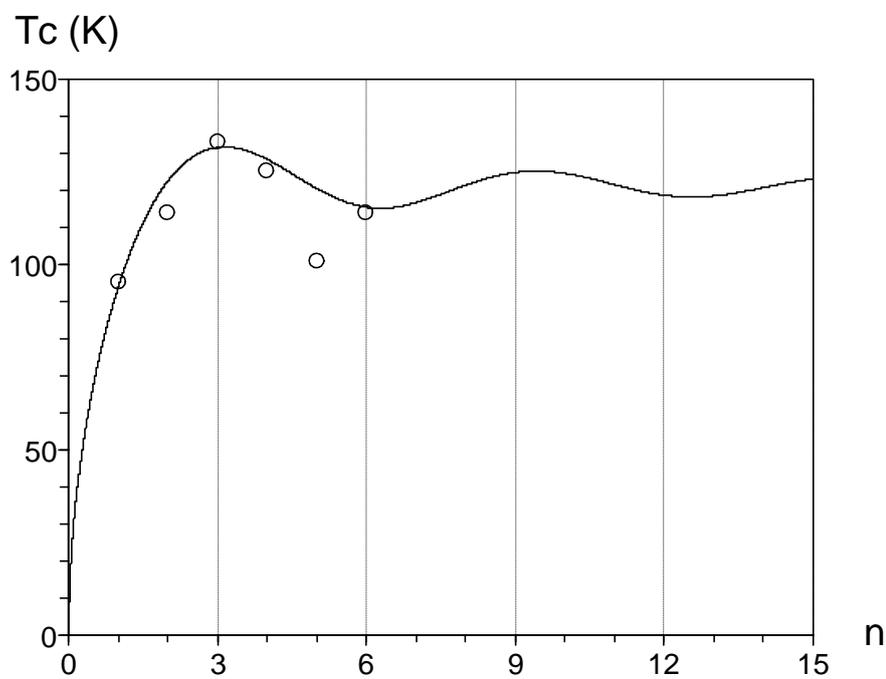

Figure 2: Plot of Tc *vs.* n from Eq.(8). $HgBa_2Ca_{n-1}Cu_nO_{2n+2}$ experimental points (○) were taken from ref. [4].

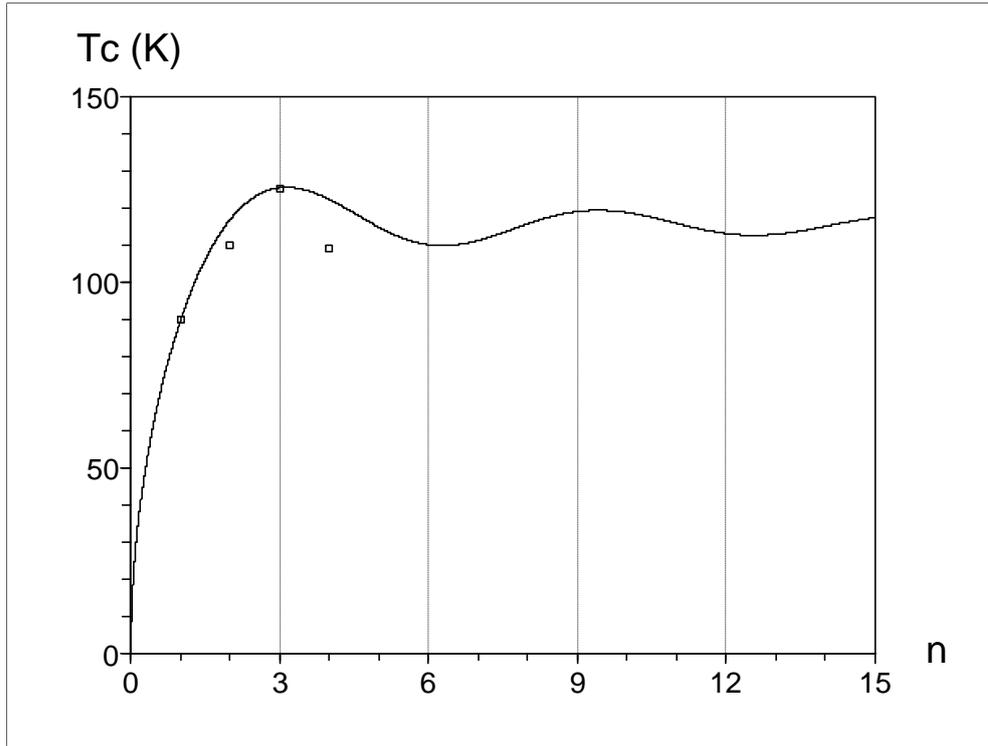

Figure 3: Plot of Tc *vs.* n from Eq.(8). $Tl_2Ba_2Ca_{n-1}Cu_nO_{2n+4}$ experimental points (□) were taken from ref. [4].

Table 1

| Family | f (n) |
|---|---|
| $TlBa_2Ca_{n-1}Cu_nO_{2n+3}$ | $\dfrac{(4n+5)^{1/3}}{(135.6n+486.9)^{1/2}}$ |
| $Bi_2Sr_2Ca_{n-1}Cu_nO_{2n+4}$ | $\dfrac{(4n+7)^{1/3}}{(135.6n+617.1)^{1/2}}$ |
| $Tl_2Ba_2Ca_{n-1}Cu_nO_{2n+4}$ | $\dfrac{(4n+7)^{1/3}}{(135.6n+707.3)^{1/2}}$ |
| $HgBa_2Ca_{n-1}Cu_nO_{2n+2}$ | $\dfrac{(4n+4)^{1/3}}{(135.6n+467.1)^{1/2}}$ |